\documentclass[preprint]{revtex4-1}
\usepackage{amsmath}
\usepackage{amssymb}
\usepackage{amsthm}
\usepackage{booktabs}
\usepackage{color}      
\usepackage{float}
\usepackage{geometry} 
\usepackage{graphicx}   
\usepackage{multirow}
\usepackage{longtable}
\usepackage{setspace}
\usepackage{subfigure}  
\usepackage{tabu}
\usepackage{verbatim}

\begin{document}
\begin{abstract}
This article explains how to apply the computer algebra package GAP (www.gap-system.org) in the computation of the problems in quantum physics, in which the application of Lie algebra is necessary. The article contains several exemplary computations which readers would follow in the desktop PC: such as, the brief review of elementary ideas of Lie algebra, the angular momentum in quantum mechanics, the quark eightfold way model, and the usage of Weyl character formula (in order to construct weight modules, and to count correctly the degeneracy).
\end{abstract}

\title{Lie algebra in quantum physics by means of computer algebra}
\author{Ichio Kikuchi}
\author{Akihito Kikuchi}
\email{akihito\_kikuchi@gakushikai.jp (The corresponding author)} 
\maketitle

\bibliographystyle{alpha}  
\section{Introduction}
In the article, titled as ``Computer Algebra and Material Design''\cite{KIKUCHI}, the author has explained how to use the computer algebra package GAP (www.gap-system.org)\cite{GAPSYSTEM} in the computation of the problems in quantum physics, especially in material physics, concerning finite groups. However, the application of the Lie algebra is omitted, although this is another important application of algebra in physics. The Lie algebra in computation by GAP, in fact, is not given by the familiar style for the physicists, but by the style of the pure mathematics, thus it is not easily applicable to the physics. So in this article, we briefly demonstrate the computation of Lie algebras in GAP, in order to digest and interpret the computed results in the language of quantum physics. As for the basics in GAP, let us consult with the article\cite{KIKUCHI}.  As for the rigorous theory of Lie algebra, see the standards text books, such as Jacobson and Humphrey\cite{JACOBSON, HUMPHR}.

\section{A short review of Lie algebra}
At first, some ideas in Lie algebra is reviewed, without proof. 

Lie algebra is an algebra generated by the commutator of the elements; the commutators also must find themselves in this algebra in the following sense: 
\begin{equation}
\left[ t_i, t_j \right ] = c_{ij}^kt_k.
\end{equation}

A simple Lie algebra is without proper ideal: we can always define subalgebra $g^{(1)}=[g,g]$, generated by commutators; if $g$ is simple, $g^{(1)}=g$.

The linear mapping $ad(a)$ is defined by $ad(a)b=\left[a,b\right]$.  The adjoint representations $ad(x)$ are by matrices $Ad(x)$ which stand for the transformation in the basis of the algebra.
\begin{equation}
ad(x)[v.1, ...., v.n]= [v.1, ...., v.n]Ad(X)
\end{equation}
From this, the Killing form is defined as $K(a,b)={\rm Tr}\left[Ad(a)Ad(b)\right]$.

We can construct a Lie algebra by a kind of basis set, given as ${e_i,f_j,h_k}$.

The maximal abelian subalgebra $H$, generated by $h_i$ is called Cartan subalgebra.
Being abelian it follows
\begin{equation}
[h_i,h_j]=0.
\end{equation} 
The Lie algebra is decomposed into the direct sum of subspaces 
\begin{equation}
L_\alpha=\{x\in L | ad(h)x=\alpha(h)x {\rm\; for\;all\;}h{\rm\; in\;}H\}, 
\end{equation}
because the Cartan algebra acts diagonally.
The eigenvalue $\alpha(h)$ is called a root, and it will be given by the entry of a vector
$\{\alpha(h_i)|i=1,...,n\}$. There is always one-to-one correspondence between an existing positive ($\alpha$) and an existing negative root ($-\alpha$). For two roots $\alpha$ and $\beta$, if $\alpha+\beta$ is a root, the commutators of the bases in the corresponding subspaces ($x_{i} \in L_{i}$) would be $[x_\alpha, x_\beta] \in L_{\alpha+\beta}$ (included in the other subspace); if not, the commutator is zero. The elements in the Cartan algebra, in this sense, generate $L_0$. We can choose the orthonormal basis ${\hat{h}_i}$ in $H$ with respect to the Killing form $K(\hat{h}_i,\hat{h}_j)=\delta_{i,j}$. The roots system is represented by the vectors in a space, now by  $(\alpha(\hat{h}_1),\alpha(\hat{h}_2),...,\alpha(\hat(h)_n))$. Here we introduce a metric other than the Killing form 
\begin{equation}
(\alpha,\beta)=\sum_i \alpha(\hat{h}_i)\beta(\hat{h}_i).
\end{equation}
We will mainly use this metric hereafter. From this geometrical image, we can introduce an ordering in the roots. We define the positive roots, when the first non-zero entry of the vector representation is positive. The simple positive roots are those which cannot be obtained by the sum of any other two positive roots. (In other words, they are the positive root vectors which are located in the edge of the convex of the positive root vectors.) Similarly the negative roots can be defined.  For convenience, we denote the basis in the Lie algebra which have the positive and negative roots by $e_i$ and $f_j$; if $i=j$, $e_i$ and $f_j$ form a pair (between $\alpha$ and $-\alpha$). That is to say, a Lie algebra could be given by the three types of the basis: \begin{equation}
\{ \{e_i|i=1,...,n\},\{f_j|j=1,...,n\},\{h_k|i=1,...,m\} \}.
\end{equation}
Now $[e_i, f_i]$  always falls upon the space $\{h_k|i=1,...,m\} \}$.

We can use representations other than that by $ad(...)$. In these cases, the eigenvalue of the elements in the Cartan subalgebra is, in general, referred as a weight. (That is to say, the weight of the adjoint representation is the root.) For example, we can use the vectors $(\omega_i(\hat{h}_1),...,\omega_i(\hat{h}_n))$ as weights, such as
\begin{equation}
\frac{2(\omega_i,\alpha_j)}{(\alpha_j,\alpha_j)}=\delta_{i,j}.
\end{equation}
This definition of weight is the fundamental weight.

Let us consider the Lie algebra generated by the angular momentum operators $J_x, J_y, J_z$. It is more convenient to represent them by the symbols $e, f, h$ as in the notations in the Lie algebra. The correspondence is given as
\begin{equation}
J_+=J_x+\sqrt{-1}J_y \rightarrow e, J_-=J_x-\sqrt{-1}J_y \rightarrow f, J_z \rightarrow h
\end{equation}
The commitators are given by
\begin{equation}
[e,f]=h, [h,e]=2e, [h,f]=-2f
\end{equation}

In quantum dynamics we denote the eigenfunction of $J_z$ as $|j,m\rangle$, with the eigenvalue m (in which we permit the half-integers). The states $|j,j\rangle...|j,-j\rangle$ are conveyed among themselves by $J_+$ and $J_l$. In Lie algebra, the eigenvalues of $h$ would be represented by integers. Now we chose a notation, where the states are denoted as $|J(=2*j),M(=2*m)):=|j,m\rangle$. The operations by the algebra are defined by
\begin{equation}
hf|J,M)=(fh-2f)|J,M)=(M-2)f|J,M),
\end{equation}
\begin{equation}
he|J,M)=(eh+2e)|J,M)=(M+2)e|J,M).
\end{equation}
The highest weight (according to the possible maximum quantum number $M$) is $|J,J)$. The other state (with lower weights) are computed by the successive operation of $e$, and the sequence $|J,J), |J,-2), ... , |J,-J)$ is finite. In general, a Lie algebra can have plural $e_i$ and $f_i$. If we put a weight as the highest one (in the sense that the operation of any $e_i$ to it leads to zero), the successive operation of $f_i$ generates the finite set of weights, as we will see later.

The $J^2=J_z^2+(J_+J_-+J_-J_+)/2$ commutes with $J_z, J_+,J_-,J_x, J_y$. It is replaced by the Casimir operator: 
$
C=[(e+f)/2]^2+[-\sqrt{-1}(e-f)/2]^2+(h/2)^2
$

The Cartan Matrix is given by $\Lambda_{i,j}=2k(\alpha_i,\alpha_j)/(\alpha_j,\alpha_j)$. This matrix enables us to utilize the graphical images of Dynkin diagrams.

\section{Composition of angular momenta}

The composition of angular momentum can be computed. We can go in two ways. The first is to use the direct product of the modules, and the second is to use the direct sum of the algebra.

Let us consider the state $|l=1,m=1\rangle|l=1/2,m=1/2\rangle$. It is more appropriate that we would denote it by $|2,2)|1,1)$, because it is a direct product of the elements of the highest weight 2 and 1 (in the sense of Lie algebra). As for the eigenstates after the composition we denote $||X,Y))$. The algebra by the Pauli matrices is the simple lie algebra A1, generated by $e,f,h$. And the universal enveloping algebra is defined now. (The enveloping algebra is the module generated by $e,f,h$ with the multiplications of the elements and the summation of the monomials generated by the multiplication.) The Lie algebra A1 is given by L, the universal enveloping algebra is UL, the generators of the latter is g.
\begin{verbatim}
gap> L:=SimpleLieAlgebra("A",1,Rationals);
<Lie algebra of dimension 3 over Rationals>
gap> UL:=UniversalEnvelopingAlgebra(L);
<algebra-with-one of dimension infinity over Rationals>
gap> g:=GeneratorsOfAlgebraWithOne(UL);
[ [(1)*x.1], [(1)*x.2], [(1)*x.3] ]
\end{verbatim}

The operator $C_2=h^2 + [e+f]^2/2 -[ e -f ]^2/2 =4ef-2h+h^2$ is twice of the so-called Casimir operator, commutes with e, f, h, and $C_2|P,Q)=P(P+2)|P,Q)$. (This operator returns four times of the eigenvalue of $J^2$ in the standard definition.) Now we define this operator in the enveloping algebra (as ``casmel'') and validate that it is in the center of the algebra. 

\begin{verbatim}
gap> casmel:=4*g[1]*g[2]-2*g[3]+g[3]*g[3];
[(4)*x.1*x.2+(-2)*x.3+(1)*x.3^2]
gap> casmel*g[1]-g[1]*casmel;
[<zero> of ...]
gap> casmel*g[2]-g[2]*casmel;
[<zero> of ...]
gap> casmel*g[3]-g[3]*casmel;
[<zero> of ...]
\end{verbatim}

The Casimir operator is redefined as a function now, so that we could apply it to the weights in the direct product of the module. The operation of the basis (v) in the Lie algebra on a weight (p) is computed by \verb!v^p!.
\begin{verbatim}
casimir:=function(p,L)
local A,B,C;
A:=(Basis(L)[2]^p);
B:=(Basis(L)[3]^p);
C:=(Basis(L)[3]^p);
A:=Basis(L)[1]^A;
C:=Basis(L)[3]^C;
return 4*A-2*B+C;
end;
\end{verbatim}

We define a module with the highest weight 1 ($|1,m),m=-1,1)$ (as V1) and another with the highest weight 2 ( $|2,m),m=-2,0,2)$ (as V2) so that the tensor product is composed (as W).

\begin{verbatim}
gap> V1:=HighestWeightModule(L,[1]);;
gap> V2:=HighestWeightModule(L,[2]);;
gap> W:=TensorProductOfAlgebraModules([V2,V1]);;
gap> Display(Basis(W));
Basis( <6-dimensional left-module over Algebra( Rationals, [ v.1, v.2, v.3 
] )>, [ 1*(1*v0<x>1*v0), 1*(1*v0<x>y1*v0), 1*(y1*v0<x>1*v0), 
 1*(y1*v0<x>y1*v0), 1*(y1^(2)*v0<x>1*v0), 1*(y1^(2)*v0<x>y1*v0) ] )
gap> V10:=HighestWeightModule(L,[10]);
<11-dimensional left-module over <Lie algebra of dimension 3 over Rationals>>
gap> Basis(V10);
Basis( <11-dimensional left-module over <Lie algebra of dimension 
3 over Rationals>>, [ 1*v0, y1*v0, y1^(2)*v0, y1^(3)*v0, y1^(4)*v0, 
  y1^(5)*v0, y1^(6)*v0, y1^(7)*v0, y1^(8)*v0, y1^(9)*v0, y1^(10)*v0 ] )
gap> List(Basis(V10),w->Basis(L)[2]^w);
[ y1*v0, 2*y1^(2)*v0, 3*y1^(3)*v0, 4*y1^(4)*v0, 5*y1^(5)*v0, 6*y1^(6)*v0, 
 7*y1^(7)*v0, 8*y1^(8)*v0, 9*y1^(9)*v0, 10*y1^(10)*v0, 0*v0 ]
gap> List(Basis(V10),w->Basis(L)[1]^w);
[ 0*v0, 10*1*v0, 9*y1*v0, 8*y1^(2)*v0, 7*y1^(3)*v0, 6*y1^(4)*v0, 5*y1^(5)*v0, 
 4*y1^(6)*v0, 3*y1^(7)*v0, 2*y1^(8)*v0, y1^(9)*v0 ]
\end{verbatim}

In the notation of GAP, the weight denoted by \verb!v0! is the highest ($|L,L)$) . The notation such as \verb!1*(y1^(p)*v0<x>y1^(q)*v0)! signifies the tensor product $\frac{f^p} {p!}|L_1,L_1) \times \frac{f^q}{q!}|L_2,L_2)$, where the denominators $p!$ and $q!$ always exist. We can see the use of this definition in the above computation, where we apply $e$ and $f$ to the basis set of the a module (V10) with the highest weight 10. The general rule of the operation to the highest weight $\lambda$ is given by
$$
v_i:=\frac{y^i}{i!}v_0, 
$$
$$
hv_i=(\lambda-2i)v_i,
ev_i=(i+1)v_{i+1},
fv_i=(\lambda-i+1)v_{i+1}.
$$

Next let us see the operations on the direct product: it is a comprehensible result if we take notice of the above rule in the notation;  the highest [\verb!(1*v0<x>1*v0)!] or the lowest [\verb!1*(1(y1)^2*v0<x>y1*v0)!] vanishes to \verb!<0-tensor>! by $e$(=Basis(L)[1]) or $f$(=Basis(L)[2]).

\begin{verbatim}
gap> List(Basis(W));
[ 1*(1*v0<x>1*v0), 1*(1*v0<x>y1*v0), 1*(y1*v0<x>1*v0), 1*(y1*v0<x>y1*v0), 
 1*(y1^(2)*v0<x>1*v0), 1*(y1^(2)*v0<x>y1*v0) ]
gap> List(Basis(W),w->Basis(L)[1]^w);
[ <0-tensor>, 1*(1*v0<x>1*v0), 2*(1*v0<x>1*v0), 
 2*(1*v0<x>y1*v0)+1*(y1*v0<x>1*v0), 1*(y1*v0<x>1*v0), 
 1*(y1*v0<x>y1*v0)+1*(y1^(2)*v0<x>1*v0) ]
gap> List(Basis(W),w->Basis(L)[2]^w);
[ 1*(1*v0<x>y1*v0)+1*(y1*v0<x>1*v0), 1*(y1*v0<x>y1*v0), 
 1*(y1*v0<x>y1*v0)+2*(y1^(2)*v0<x>1*v0), 2*(y1^(2)*v0<x>y1*v0), 
 1*(y1^(2)*v0<x>y1*v0), <0-tensor> ]
\end{verbatim}

Let us decompose the vector space of the tensor products, according to the eigenvalue of the Casimir operator (to be exact, the twice of the proper definition.) This space has two irreducible subspaces, which have the eigenvalues 15 and 3, i.e. $||3,m))$ and $||1,m))$. 
The operation of the Casimir operator is stored in the list CASM, and represented by a matrix MAT, so that we compute the eigenvalues and the eigenvectors at EIGV and EIGS:  
\begin{verbatim}
gap> CASM:=List(Basis(W),w->casimir(w,L));
[ 15*(1*v0<x>1*v0), 7*(1*v0<x>y1*v0)+4*(y1*v0<x>1*v0), 
 8*(1*v0<x>y1*v0)+11*(y1*v0<x>1*v0), 11*(y1*v0<x>y1*v0)+8*(y1^(2)*v0<x>1*v0),
 4*(y1*v0<x>y1*v0)+7*(y1^(2)*v0<x>1*v0), 15*(y1^(2)*v0<x>y1*v0) ]
gap> MAT:=List(CASM,c->Coefficients(Basis(W),c));
[ [ 15, 0, 0, 0, 0, 0 ], [ 0, 7, 4, 0, 0, 0 ], [ 0, 8, 11, 0, 0, 0 ], 
 [ 0, 0, 0, 11, 8, 0 ], [ 0, 0, 0, 4, 7, 0 ], [ 0, 0, 0, 0, 0, 15 ] ]
gap> EIGV:=Eigenvalues(Rationals,MAT);;
gap> EIGS:=Eigenspaces(Rationals,MAT);;
gap> Display(EIGV);
[ 15, 3 ]
gap> Display(EIGS);
[ VectorSpace( Rationals, [ [ 1, 0, 0, 0, 0, 0 ], [ 0, 1, 1, 0, 0, 0 ], 
     [ 0, 0, 0, 1, 1, 0 ], [ 0, 0, 0, 0, 0, 1 ] ] ), 
 VectorSpace( Rationals, [ [ 0, 1, -1/2, 0, 0, 0 ], [ 0, 0, 0, 1, -2, 0 ] 
    ] ) ]
\end{verbatim}

The list EIGS include two irreducible vector spaces, accessible as EIGS[1] and EIGS[2], for eigenvalue 15 and 3. (As we see, each of them shows us the eigenvectors.) Let us analyze the eigenspaces. In the following, the list EV15 includes the eigenvectors of the eigenvalue 15, represented in the weight module,  
to which the elements $\{e,f,h\}$ =$\{$Basis(L)[1],Basis(L)[2],Basis(L)[3]$\}$ are operated.
We see that the operations by $e$ and $f$ shift the states in this irreducible space (in the second and the third computation) and that the elements $h$ acts by the eigenvalues 3, 1, -1, -3 by turns (in the fourth computation), as we expect.

\begin{verbatim}
gap> EV15:=List(Basis(EIGS[1]),b->b*Basis(W));
[ 1*(1*v0<x>1*v0), 1*(1*v0<x>y1*v0)+1*(y1*v0<x>1*v0), 
  1*(y1*v0<x>y1*v0)+1*(y1^(2)*v0<x>1*v0), 1*(y1^(2)*v0<x>y1*v0) ]
gap> List(EV15, v->Basis(L)[1]^v);
[ <0-tensor>, 3*(1*v0<x>1*v0), 2*(1*v0<x>y1*v0)+2*(y1*v0<x>1*v0), 
  1*(y1*v0<x>y1*v0)+1*(y1^(2)*v0<x>1*v0) ]
gap> List(EV15, v->Basis(L)[2]^v);
[ 1*(1*v0<x>y1*v0)+1*(y1*v0<x>1*v0), 
  2*(y1*v0<x>y1*v0)+2*(y1^(2)*v0<x>1*v0), 
  3*(y1^(2)*v0<x>y1*v0), <0-tensor> ]
gap> List(EV15, v->Basis(L)[3]^v);
[ 3*(1*v0<x>1*v0), 1*(1*v0<x>y1*v0)+1*(y1*v0<x>1*v0), 
  -1*(y1*v0<x>y1*v0)-1*(y1^(2)*v0<x>1*v0), -3*(y1^(2)*v0<x>y1*v0) ]
\end{verbatim}

The result is summarized as follows. (The bases are numbered as No.1,2,3,4.)

\begin{verbatim}
       No.1  ||3,3))               No.2 ||3,1))               
       1*(1*v0<x>1*v0)             1*(1*v0<x>y1*v0)
                                   +1*(y1*v0<x>1*v0)    
by e    goes to zero               goes to No.1        
by f    goes to No.2               goes to No.3        
by h    gets eigenvalue 3          gets eigenvalue 1        

        No.3 ||3,-1))              No.4 ||3,-3))
        1*(y1*v0<x>y1*v0)          1*(y1^(2)*v0<x>y1*v0)     
        +1*(1*(y1^(2)*v0<x>1*v0) 
by e    goes to No.2               goes to No.4
by f    goes to No.4               goes to zero
by h    gets eigenvalue -1         gets eigenvalue -3
\end{verbatim}

In fact, although the bases in the weight module are assumed to be orthogonal, they are not normalized in the necessary way of quantum physics, so as to be $(x|x)=1$, because the weights are generated by the operation $f^n/n!|J,J)$. Thus the states $||3,1))=[ 0, 1, 1, 0, 0, 0 ]$ and $||1,1))=[ 0, 1, -1/2, 0, 0, 0 ]$ (and also another pair) falsely seem to be not orthogonal. We must normalize them if necessary, according to $e|J,M)=\sqrt{(J-M)(J+M+1)}/2|J,M+2)$ and $f|J,M)=\sqrt{(J+M)(J-M+1)}/2|J,M-2)$. 

We can also consider in another way, by the direct sum of the algebra $L_1 + L_2$ : the components $L_1$ and $L_2$ act upon the direct product of modules. The direct sum is given in K, and the highest weight module is V, constructed from two module with the highest weight 2 and 1. The Casimir operator is again defined by a function casimir2, and its operation of the module is stored in CASM2 and MAT2 so that the eigenvalues and the eigenvectors are computed at EIGV2 and EIGS2.   

\begin{verbatim}
gap> L:=SimpleLieAlgebra("A",1,Rationals);;
gap> K:=DirectSumOfAlgebras([L,L]);;
gap> V:=HighestWeightModule(K,[2,1]);; 
gap> Display(Basis(V));
Basis( <6-dimensional left-module over Algebra( Rationals, 
[ v.1, v.2, v.3, v.4, v.5, v.6 ] )>, [ 1*v0, y1*v0, y2*v0, y1^(2)*v0, 
 y1*y2*v0, y1^(2)*y2*v0 ] )
\end{verbatim}
The basis in the direct sum of algebras is given by
\begin{verbatim}
[ v.1, v.2, v.3, v.4, v.5, v.6 ],
\end{verbatim}
and that in the module is given by
\begin{verbatim}
[ 1*v0, y1*v0, y2*v0, y1^(2)*v0, y1*y2*v0, y1^(2)*y2*v0 ]
\end{verbatim}
where \verb![ v.1, v.2, v.3 ]! and \verb![ v.4, v.5, v.6 ]! stand for $[e_1,f_1,h_1]$ and$[e_2,f_2,h_2]$ respectively for two Lie algebra $L_1$ and $L_2$.  Now the operations by $f_1$ and $f_2$ are denoted by \verb!y1*v0! and \verb!y2*v0!, and the weights in the modules are not explicitly denoted by the tensor products. If we keep in mind these notations, however, the computation goes in the similar way as in the previous example.

\begin{verbatim}
gap> casimir2:=function(p,K)
local A,B,C;
A:=(Basis(K)[2]^p)+(Basis(K)[5]^p);
B:=(Basis(K)[3]^p)+(Basis(K)[6]^p);
C:=(Basis(K)[3]^p)+(Basis(K)[6]^p);
A:=Basis(K)[1]^A+(Basis(K)[4]^A);
C:=Basis(K)[3]^C+(Basis(K)[6]^C);
return 4*A-2*B+C;
end;

gap> CASM2:=List(Basis(V),w->casimir2(w,K));
[ 15*1*v0, 11*y1*v0+8*y2*v0, 4*y1*v0+7*y2*v0, 7*y1^(2)*v0+4*y1*y2*v0, 
 8*y1^(2)*v0+11*y1*y2*v0, 15*y1^(2)*y2*v0 ]
gap> MAT2:=List(CASM2,c->Coefficients(Basis(V),c));
[ [ 15, 0, 0, 0, 0, 0 ], [ 0, 11, 8, 0, 0, 0 ], [ 0, 4, 7, 0, 0, 0 ], 
 [ 0, 0, 0, 7, 4, 0 ], [ 0, 0, 0, 8, 11, 0 ], [ 0, 0, 0, 0, 0, 15 ] ]
gap> EIGV2:=Eigenvalues(Rationals,MAT2);;
gap> EIGS2:=Eigenspaces(Rationals,MAT2);;
gap>  Display(EIGV2);
[ 15, 3 ]
gap> Display(EIGS2);
[ VectorSpace( Rationals, [ [ 1, 0, 0, 0, 0, 0 ], [ 0, 1, 1, 0, 0, 0 ], 
     [ 0, 0, 0, 1, 1, 0 ], [ 0, 0, 0, 0, 0, 1 ] ] ), 
 VectorSpace( Rationals, [ [ 0, 1, -2, 0, 0, 0 ], [ 0, 0, 0, 1, -1/2, 0 ] 
    ] ) ]
\end{verbatim}

\section{Quark model}
\begin{figure}[H]
\includegraphics[width=1\linewidth]{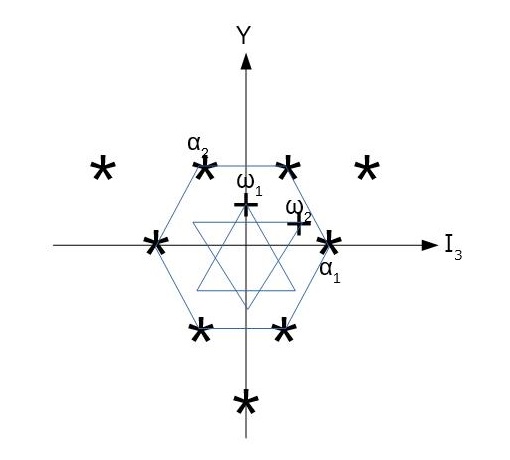}
\caption{The roots and the fundamental weights are illustrated in the plane. Two simple positive roots are denoted by $\alpha_1$ and $\alpha_2$. The fundamental weights are denoted by $\omega_1$ and $\omega_2$. The module of the quarks is given by the points in the central inverted triangle, which includes $\omega_1$ as the highest weight. The antiquarks is another central triangle, which includes $\omega_2$ as the highest weight. The module of the octet baryon is given by the six dots in the hexagon and the origin with the duplication, where the highest root is $\alpha_1+\alpha_2$. The decuplet is composed by 10 elements, which are the green and the red dots and the origin, located in the large inverted triangle, where the highest weight is the upper right vertex.
}
\label{ROOTSA2}
\end{figure}

In this section, the quantum states of hadrons are classified according to the symmetry of $su(3)$, which has another name $A_2$. This algebra has two primitive roots $\alpha_1$ and $\alpha_2$, and we put them in a plane. If one uses the metric $(\alpha,\beta)$, defined in the previous section, these roots $\alpha_i$ and the fundamental weights $\omega_i$ can be illustrated in the points in the triangular lattices, as in Fig.\ref{ROOTSA2}.

We can classify the symmetry of the hadrons through the idea of the modules over semi-simple Lie algebras, which have the highest weights, with respect to the fundamental weights. (The construction of the module from the highest weight is briefly given in the next section.) 

According to the double index of the highest weights, the quark multiplets are given as
\begin{itemize}
\item Quark triplet ${\bf 3}=[1,0]$, dimension 3,
\item Antiquark triplet ${\bf 3}^*=[0,1]$, dimension 3,
\item Octet baryon triplet ${\bf 8}=[1,1]$, dimension 8,
\item Decuplet baryon triplet ${\bf 10}=[3,0]$, dimension 10.
\end{itemize}

Let us start the computation. At first we derive elementary information. 
\begin{verbatim}
gap> K:=SimpleLieAlgebra("A",2,Rationals);
<Lie algebra of dimension 8 over Rationals>
gap> R:=RootSystem(K);
<root system of rank 2>
gap> CartanMatrix(R);
[ [ 2, -1 ], [ -1, 2 ] ]
gap> BilinearFormMat( R )
[ [ 1/3, -1/6 ], [ -1/6, 1/3 ] ]
gap> PositiveRootVectors(R);
[ v.1, v.2, v.3 ]
gap> PositiveRoots(R);
[ [ 2, -1 ], [ -1, 2 ], [ 1, 1 ] ]
gap> NegativeRootVectors(R);
[ v.4, v.5, v.6 ]
gap> NegativeRoots(R);
[ [ -2, 1 ], [ 1, -2 ], [ -1, -1 ] ]
gap> CB:=ChevalleyBasis(K);
[ [ v.1, v.2, v.3 ], [ v.4, v.5, v.6 ], [ v.7, v.8 ] ]
gap> CB[3]*CB[1];
(2)*v.1+(2)*v.2
gap> CB[1];
[ v.1, v.2, v.3 ]
gap> CB[3][1]*CB[1];
[ (2)*v.1, (-1)*v.2, v.3 ]
gap> CB[3][2]*CB[1];
[ (-1)*v.1, (2)*v.2, v.3 ]
gap> CB[3][2]*CB[2];
[ v.4, (-2)*v.5, (-1)*v.6 ]
gap> CB[3][2]*CB[1];
[ (-1)*v.1, (2)*v.2, v.3 ]
gap> CB[3][2]*CB[2];
[ v.4, (-2)*v.5, (-1)*v.6 ]
gap> KB:=KillingMatrix(Basis(K));
[ [ 0, 0, 0, 6, 0, 0, 0, 0 ], [ 0, 0, 0, 0, 6, 0, 0, 0 ], 
 [ 0, 0, 0, 0, 0, 6, 0, 0 ], [ 6, 0, 0, 0, 0, 0, 0, 0 ], 
 [ 0, 6, 0, 0, 0, 0, 0, 0 ], [ 0, 0, 6, 0, 0, 0, 0, 0 ], 
 [ 0, 0, 0, 0, 0, 0, 12, -6 ], [ 0, 0, 0, 0, 0, 0, -6, 12] ]
\end{verbatim}

The command \verb!BilinearFormMat()! returns $(\alpha_i,\alpha_j)$ for simple positive roots. From the result, we see that there are two simple positive roots, and they form the angle of 120 degrees if they are put on the plane. There are eight bases in A2; \verb!v.1 v.2 v.3! are the positive roots; \verb!v.4 v.5 v.6! are the negative roots; \verb!v.7 and v.8! are the elements of Cartan subalgebra. The command ``Chevallaybasis'' returns the basis in this sorting. We see the operations of the Cartan algebra on the roots (simply done by the product \verb!*!) give the labels of the roots through the eigenvalues. The Killing matrix is computed in the last, from which we see that the elements of Cartan subalgebra are not orthonormalized. But $K(h_i,h_j)$ is, in this case, \verb![[12, -6],[-6,12]]! (in the lower right corner of the matrix), so we can compute $(\alpha_i,\alpha_j)$ as follows.
\begin{verbatim}
gap> [2,-1]*([[12,-6],[-6,12]]^-1)*[2,-1];
1/3
gap> [2,-1]*([[12,-6],[-6,12]]^-1)*[-1,2];
-1/6
\end{verbatim}

Now we compute the modules of the quark multiplets by the maximal weight [1,0] (as Q) , the antiquarks by the [0,1] (as AQ), the octet by [1,1] (as OB) , the decuplet by [3,0] (as DB).
\begin{verbatim}
gap> Q:=HighestWeightModule(K,[1,0]); 
<3-dimensional left-module over <Lie algebra of dimension 8 over Rationals>>
gap> AQ:=HighestWeightModule(K,[0,1]);
<3-dimensional left-module over <Lie algebra of dimension 8 over Rationals>>
gap> OB:=HighestWeightModule(K,[1,1]);
<8-dimensional left-module over <Lie algebra of dimension 8 over Rationals>>
gap> DB:=HighestWeightModule(K,[3,0]);
<10-dimensional left-module over <Lie algebra of dimension 8 over Rationals>>
\end{verbatim}

\begin{figure}[H]
\includegraphics[width=1\linewidth]{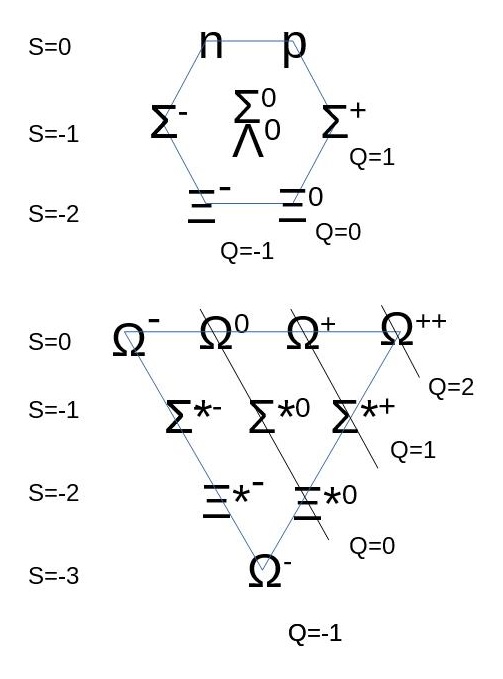}
\caption{The diagrams of the octet (above) and the decuplet (below) are shown. }
\label{BARYONS}
\end{figure}

The elements in these quantum states are given in Fig.\ref{ROOTSA2}, in the coordinate system with respect to the quantum number $I_3$ (isospin) and $Y$ (hypercharge). The baryon octet and decuplet are also given in the Fig. \ref{BARYONS}, in the same coordinate system, with quantum numbers $S$ and $Q$.  In this figure, the corresponding baryons are put over the weights of the modules. The quantum number spin $S$ is defined by $S=Y-B$, where $B=1$ the baryonic number. $Q=I_3+Y/2$ is the electronic charge, and the baryons which have the same $Q$ are connected by oblique lines.

Here I will show how to compare the computed result with the above figures of the triplet, octet, and decuplet representations. The information on the bases of the module is obtained in this way:
\begin{verbatim}
gap> List(Basis(DB),e->ExtRepOfObj(e));
[ 1*v0, y1*v0, y3*v0, y1^(2)*v0, y1*y3*v0, y1^(3)*v0, y3^(2)*v0, 
 y1^(2)*y3*v0, y1*y3^(2)*v0, y3^(3)*v0 ]
gap> List(Basis(DB),e->ExtRepOfObj(ExtRepOfObj(e)));
[ [ [ 1, 1, [ 3, 0 ] ], 1 ], [ [ 2, y1, [ 1, 1 ] ], 1 ], 
 [ [ 3, y3, [ 2, -1 ] ], 1 ], [ [ 4, y1^(2), [ -1, 2 ] ], 1 ], 
 [ [ 5, y1*y3, [ 0, 0 ] ], 1 ], [ [ 6, y1^(3), [ -3, 3 ] ], 1 ], 
 [ [ 7, y3^(2), [ 1, -2 ] ], 1 ], [ [ 8, y1^(2)*y3, [ -2, 1 ] ], 1 ], 
 [ [ 9, y1*y3^(2), [ -1, -1 ] ], 1 ], [ [ 10, y3^(3), [ 0, -3 ] ], 1 ] ]
\end{verbatim}

The operations of y1( by v.4), y2( by v.5), and y3( by v.6) are successively executed on the highest weight \verb!v0! and the weights are generated. The computed results are returned by the list, such as \verb![ 9, y1*y3^(2), [ -1, -1 ] ], 1 ]!. The first entry is the sequential number; the second entry, \verb!y1*y3^(2)! signifies the operations of y1, y2, or y3 to the highest weight \verb!v0!; the third entry, \verb![-1, -1]!,  is the coordinate with respect to the fundamental weight, and the corresponding weight is given by $-1\cdot\omega_1-1\cdot\omega_2$. Now the weights in the module of quarks, antiquarks, octet, and decuplet are listed in the below.  

\begin{verbatim}
Quarks
 [ [ 1, 1, [ 1, 0 ] ], 1 ]
 [ [ 2, y1, [ -1, 1 ] ], 1 ] 
 [ [ 3, y3, [ 0, -1 ] ], 1 ] 

Antiquarks
 [ [ 1, 1, [ 0, 1 ] ], 1 ] 
 [ [ 2, y2, [ 1, -1 ] ], 1 ] 
 [ [ 3, y3, [ -1, 0 ] ], 1 ] 

Baryon Octet
 [ [ 1, 1, [ 1, 1 ] ], 1 ]               p+
 [ [ 2, y1, [ -1, 2 ] ], 1 ]             n0
 [ [ 3, y2, [ 2, -1 ] ], 1 ]             SIGMA+
 [ [ 4, y1*y2, [ 0, 0 ] ], 1 ]           LAMBDA0 or SIGMA0
 [ [ 5, y3, [ 0, 0 ] ], 1 ]              SIGMA0  or LAMBDA0
 [ [ 6, y1*y3, [ -2, 1 ] ], 1 ]          SIGMA-
 [ [ 7, y2*y3, [ 1, -2 ] ], 1 ]          XI0
 [ [ 8, y3^(2), [ -1, -1 ] ], 1 ]        XI-

Baryon decuplet
 [ [ 1, 1, [ 3, 0 ] ], 1 ]               DELTA++
 [ [ 2, y1, [ 1, 1 ] ], 1 ]              DELTA+
 [ [ 3, y3, [ 2, -1 ] ], 1 ]             SIGMA*+ 
 [ [ 4, y1^(2), [ -1, 2 ] ], 1 ]         DELTA0 
 [ [ 5, y1*y3, [ 0, 0 ] ], 1 ]           SIGMA*0
 [ [ 6, y1^(3), [ -3, 3 ] ], 1 ]         DELTA-
 [ [ 7, y3^(2), [ 1, -2 ] ], 1 ]         XI*0
 [ [ 8, y1^(2)*y3, [ -2, 1 ] ], 1 ]      SIGMA*-
 [ [ 9, y1*y3^(2), [ -1, -1 ] ], 1 ]     XI*-
 [ [ 10, y3^(3), [ 0, -3 ] ], 1 ]        OMEGA-
\end{verbatim}

Let us apply \verb!v.4(=y1)!, \verb!v.5(=y2)! and \verb!v.6(=y3)! to the elements in the modules of the decuplet.
\begin{verbatim}
gap> List(Basis(DB));
[ 1*v0, y1*v0, y3*v0, y1^(2)*v0, y1*y3*v0, y1^(3)*v0, y3^(2)*v0, 
 y1^(2)*y3*v0, y1*y3^(2)*v0, y3^(3)*v0 ]
gap> List(Basis(DB),w->Basis(K)[4]^w);
[ y1*v0, 2*y1^(2)*v0, y1*y3*v0, 3*y1^(3)*v0, 2*y1^(2)*y3*v0, 0*v0, 
 y1*y3^(2)*v0, 0*v0, 0*v0, 0*v0 ]
gap> List(Basis(DB),w->Basis(K)[5]^w);
[ 0*v0, -1*y3*v0, 0*v0, -1*y1*y3*v0, -2*y3^(2)*v0, -1*y1^(2)*y3*v0, 
 0*v0, -2*y1*y3^(2)*v0, -3*y3^(3)*v0, 0*v0 ]
gap> List(Basis(DB),w->Basis(K)[6]^w);
[ y3*v0, y1*y3*v0, 2*y3^(2)*v0, y1^(2)*y3*v0, 2*y1*y3^(2)*v0, 0*v0, 
 3*y3^(3)*v0, 0*v0, 0*v0, 0*v0 ]
\end{verbatim}

The result is arranged on the table in the below. In the leftmost column, the elements in the decuplet are given. In the other columns, the result of the operations by y1 y2 y3 denoted by \verb!y1^ y2^ y3^! (according to the notation of GAP) is shown. We can see that the operations of y1, y2 and y3 move the weights, by $-\alpha_1$, $-\alpha_1$, $-\alpha_1-\alpha_2$: if a basis would go outside of the triangle of the decuplet, it becomes zero, \verb!0*v0!. We can also check that the operations by x1, x2, and x3 cause the shifts of the positions in the inverse directions. 

\begin{verbatim} 
                  y1^             y2^              y3^ 
[ [ 1*v0,         y1*v0,          0*v0,            y3*v0 ], 
  [ y1*v0,        2*y1^(2)*v0,    -1*y3*v0,        y1*y3*v0 ], 
  [ y3*v0,        y1*y3*v0,       0*v0,            2*y3^(2)*v0 ], 
  [ y1^(2)*v0,    3*y1^(3)*v0,    -1*y1*y3*v0,     y1^(2)*y3*v0 ], 
  [ y1*y3*v0,     2*y1^(2)*y3*v0, -2*y3^(2)*v0,    2*y1*y3^(2)*v0 ], 
  [ y1^(3)*v0,    0*v0,           -1*y1^(2)*y3*v0, 0*v0 ], 
  [ y3^(2)*v0,    y1*y3^(2)*v0,   0*v0,            3*y3^(3)*v0 ], 
  [ y1^(2)*y3*v0, 0*v0,           -2*y1*y3^(2)*v0, 0*v0 ], 
  [ y1*y3^(2)*v0, 0*v0,           -3*y3^(3)*v0,    0*v0 ], 
  [ y3^(3)*v0,    0*v0,           0*v0,            0*v0 ]  ]
\end{verbatim}

It is not trivial why the representations of the weights take the particular geometrical shapes, in some cases small or large triangles, in another, a hexagon. We will inspect this question in the next section.

\section{Highest weight construction}
The process of obtaining all weights from the highest weight is briefly reviewed. If we apply $e_i$ and $f_i$ to a weight $v$ we obtain this succession:
$$
f_i^qv,f_i^{q-1}v...,f_iv,v,e_iv,...,e_i^pv.
$$

The operation by $e_i$ and $f_i$ make a shift in the weight, from $\mu$ to $\mu+\alpha_i$ and $\mu-\alpha_i$, by a simple positive root $\alpha_i$. The succession is bound by two integers $p$ and $q$ by the finiteness. We have a relation:
$$
\frac{2(v,\alpha_i)}{(\alpha_i,\alpha_i)}=-(p-q).
$$
(This relation is derived from the argument on the Weyl reflection.)

The highest weight $\mu$ is a weight if $\mu+\alpha$ is not a weight for any positive root $\alpha$. If we start from the highest weight, it happens that $p=0$, so that we have a succession:
$$
f_i^qv, f_i^{q-1}v,...,f_iv,v.
$$
In general there are plural simple positive roots. So if simple positive roots $\alpha_i$, by the operations $e_i$, permit a highest weight (which should be represented by the fundamental weights) to go down $q_1$, $q_2$,...,$q_n$ times in each direction of $-\alpha_i$, we can denote this highest weight by the label [$q_1$,$q_2$ ,...,$q_n$]. The other weights are likewise labeled by $-2(v,\alpha_l)(\alpha_l,\alpha_l)$, that is, [$q_1-p_1$, $q_2-p_2$ ,...,$q_n-p_n$]. The shift along $-\alpha_j$ (from $x$ to $f_j x$) causes the change at the i-th entry of the label by $-2(\alpha_j,\alpha_i)/(\alpha_i,\alpha_i)$. (This is nothing but the subtraction of vector, that of a root from a weight, provided that the vectors are given in the coordinate system by the fundamental weights.)

Let us consider the highest weight $[1,1]$ in the simple Lie algebra $su(3)$ (A2).  There are two positive simple roots $\alpha_1$ and $\alpha_2$, and the changes in the label by these two shift are given by $-[2,-1]$ and $-[-1,2]$. The highest weight $[1,1]$ can move along $-\alpha_1$ and $-\alpha_2$, and gives birth to two weights $[-1,2]$ and $[2,-1]$. In the former, $[-1,2]=f_1*[1,1]$, so $(p_1,p_2)=(1,0)$. (This weight could make one movement by $e_1$ toward the highest weight, but from it, no more.) The values $p_i-q_i$ are two entries in the brackets: $(q_1-p_1,q_2-p_2)=[-1,2]$, so $(q_1,q_2)=(0,2)$. Then the weight $[-1,2]$ can shift twice along $-\alpha_2$ direction, and gives birth to $[0,0]$ and $[1,-2]$. Similarly the weight $[2,-1]$ shifts twice along $-\alpha_1$ direction and gives birth to $[0,0]$ and $[-2,1]$. Finally, the weights $[-2,1]$ and $[1,-2]$ reach $[-1,-1]$. (As for $[1,-2]$, because $(p_1,p_2)=(2,1)$ and $(p_1-q_1,p_2-q_2)=(1,-2)$, so $(q_1,q_2)=(1,0)$; it descends only once along $-\alpha_1$.) We obtain seven labels, but in fact, eight weights, because $[0,0]$ has double degeneracy. This is because this [1,1]-representation is equivalent to the adjoint representation of A2, where the elements in the algebra play the roles of the operators and the bases in the module at the same time. In the adjoint representation, there are three positive roots (e1,e2,e3),three negative roots (f1,f2,f3), and two Cartan generators (h1,h2); the correspondence is given by e1$\rightarrow$[2,-1], e2$\rightarrow$[-1,2], e3$\rightarrow$[1,1],f1$\rightarrow$[-2,1], f2$\rightarrow$[1,-2], e3$\rightarrow$[-1,-1], h1$\rightarrow$[0,0], h2$\rightarrow$[0,0]. (However, this two hold degeneracy at [0,0] does not always happen, although the set of the weights might include the hexagon in [1,1]-representation, such as in the case of [3,0]-representation, shown in the previous section.) 

The evaluation of degeneracy of the weights requires more advanced knowledge, concerning the Weyl character formula. This formula is symbolically written as
$$
\chi_\lambda=\frac{\sum_{w \in W} (-1)^w \exp(w(\lambda+\rho))}{\exp(\rho)\prod_{\alpha \in \Delta^+}(1-\exp(-\alpha))}
$$
where $\lambda$ is a highest weight; $W$ is the Weyl group (a group generated by the reflection of a weight $l$ with respect to the hyperplane perpendicular to the simple root $\alpha_i$, that is, $w(l):=l -2(l,\alpha_i)/(\alpha_i,\alpha_i) )$; $\Delta^+$ is the set of the positive roots and $\rho=\sum_{\alpha \in \Delta^+}\alpha/2$; $(-1)^w$ stands for the odd or even parity of the reflection.

Let us try the formula.

We define the Lie algebra A2 (as K), the root system (as R), and the Weyl group(as W). The elements in W is stored in list ELW, and $\rho$ is in RHO.
\begin{verbatim}
gap> K:=SimpleLieAlgebra("A",2,Rationals);;
gap> R:=RootSystem(K);;
gap> W:=WeylGroup(R);;
gap> ELW:=Elements(W);;
gap> RHO:=Sum(PositiveRoots(R))/2;;
\end{verbatim}

We use the highest weight $\lambda$=[1,1], and the Weyl reflections on $\lambda+\rho$ are stored in a list (RFW), where the operation of the Weyl group is computed by the multiplication of a row vector (in left) and a matrix (in right).  The parity $(-1)^w$ in the formula is simply computed by the determinant of the matrix representation of $w$.

\begin{verbatim}
gap> AWEIGHT:=[1,1];;
gap> RFW:=List(ELW,e->[Determinant(e),(AWEIGHT+RHO)*e]);;
\end{verbatim}

To construct the formula, we use the variables x, y, which would stand for $\exp(\omega_1)$ and $\exp(\omega_2)$ with respect to two fundamental weights $\omega_1$ and $\omega_2$. The denominators and the numerators are computed at NUMERATOR and DENOMINATOR, and the character is at WEYLCHR.

\begin{verbatim}
gap> x:=Indeterminate(Rationals,"x");;
gap> y:=Indeterminate(Rationals,"y");;
gap> DENOMINATOR:=x^RHO[1]*y^RHO[2]
    *Product(List(PositiveRoots(R),i->1-x^(-i[1])*y^(-i[2])));
(x^3*y^3-x^4*y-x*y^4+x^3+y^3-x*y)/(x^2*y^2)
gap> NUMERATOR:=Sum(List(RFW, e-> e[1]*x^e[2][1]*y^e[2][2]));
(-x^8*y^6+x^10*y^2+x^4*y^8-x^8-x^2*y^6+x^4*y^2)/(-x^6*y^4)
gap>WEYLCHR:=NUMERATOR/DENOMINATOR;
(-x^5*y^3-x^6*y-x^3*y^4-2*x^4*y^2-x^5-x^2*y^3-x^3*y)/(-x^4*y^2)
\end{verbatim}

The result is returned by a rational function, but it is actually a polynomial. Let us analyze it. The list of the monomials, coefficients in the polynomial, and the corresponding weights are now given:
\begin{verbatim}
 Monomial     Coefficient     Weight
 x^1*y^1         1            [ 1,  1]
 x^2*y^(-1)      1            [ 2, -1]
 x^(-1)*y^2      1            [-1,  2]
 x^0*y^0         2            [ 0,  0]
 x^1*y^(-2)      1            [ 1, -2]
 x^(-1)*y^1      1            [-1,  1]
 x^(-1)*y^(-1)   1            [-1, -1]
\end{verbatim}
The monomial which represents the weight [0,0] has the coefficient 2, and the others have 1. That is to say, the generated monomials in the formula stand for the weights in the module, and the coefficient to monomials give the degeneracy of the weights.


Finally, an exercise is left to the readers. to construct the direct product of three quarks through the computations given in this and previous section.
\begin{itemize}
\item Let us define the quark triplet.
\item Let us compose the direct product of three triplets.
\item Let us find out the definition of general secondary Casimir operator (  usually given by means of the Killing matrix).
\item Let us apply the Casimir operator to the basis set in the direct product of the triplets. 
\item Let us analyze the operation of the Casimir operator as an eigenvalue problem.
\item We will find three types of eigenspaces: the first, with one dimension; the second, with 16 dimensions; the third, with 10 dimensions. The second includes the two copies of the baryon octets, and the third the baryon decuplet.
\item In this wise, we can obtain explicit representations of baryons without the usage of Young tableaux. 
\end{itemize}

\section{summary}
In this article, the computation by GAP concerning Lie algebra physics is demonstrated. The computed results by GAP would be returned in a somewhat brusque way, so the authors have explained how to interpret them according to the context of quantum physics, with the hope that this short note would be of use to the physicists and the students who would like to apply the computer algebra in the actual purpose.

\end{document}